\documentstyle[12pt]{article}
\def\beq{\begin{equation}}   \def\eeq{\end{equation}}
\def\bea{\begin{eqnarray}}   \def\eea{\end{eqnarray}}

\begin{document}

\thispagestyle{empty}
\begin{titlepage}

\begin{flushright}
DESY 00-183 \\
TTP 00--28 \\
UND-HEP-00-BIG12 \\
hep-ph/0012218
\end{flushright}

\vspace{0.3cm}
\boldmath
\begin{center}
\Large\bf $b$ Quark Physics with $2 \times 10^9$ $Z$ Bosons
\end{center}
\unboldmath
\vspace{0.8cm}

\begin{center}
{\large Ahmed Ali} \\
{\sl DESY Theory Group, Notkestrasse 85, 22607 Hamburg, Germany}
\vspace*{2mm} \\
{\large Donald Benson},
{\large Ikaros Bigi} \\
{\sl Physics Dept., Univ. of Notre Dame du Lac,
     Notre Dame, IN 46556, U.S.A}
\vspace*{2mm} \\
{\large Richard Hawkings}\\
{\sl CERN--EP, CH-1211 Geneva, Switzerland}
\vspace*{2mm} \\
{\large Thomas Mannel}\footnote{On leave from
{\sl Institut f\"{u}r Theoretische Teilchenphysik, \\
     D -- 76128 Karlsruhe, Germany.}} \\
{\sl CERN--TH, CH-1211 Geneva 23, Switzerland }
\end{center}

\vspace{\fill}

\begin{abstract}
\noindent
It has been suggested to realize a factory for
$10^9$ $Z^0$
through a linear $e^+e^-$ collider with polarized beams.
Very likely the relevant $CP$ studies for $B$ mesons
will already have been
performed at the $B$ factories by that time,
hence GIGA-Z  will be a
third generation $b$ physics experiment.
Yet such a facility would provide us
with unique opportunities in the domain of
beauty physics that would be of essential significance
even in 2010: (1) Production and decay of polarized beauty
baryons; (2) searching for and probing transitions driven by
$b \to q \nu \bar \nu$; (3) detailed and comprehensive
studies of inclusive semileptonic $B_s$ decays.
\end{abstract}
\end{titlepage}
\section{GIGA-Z vs. its competition: LHC-b/BTeV and BELLE/BaBar}
A high-luminosity linear collider running on the $Z^0$ resonance
offers intriguing and even unique possibilities for $b$ quark physics.
Assuming a sample of $2 \times 10^9$ $Z^0$ bosons produced per year
at such a facility, one would end up with about $6 \times 10^8$ $b$
or $\bar{b}$ quarks produced in a relatively clean environment.
Furthermore, the option of polarized beams would create
exciting capabilities
for polarization studies and greatly improved $b$-flavour tagging.

Given the time scales of the linear collider projects $b$ physics
studies at GIGA-Z would be a third generation experiment.
Thus in evaluating the $b$-physics potential of this
possibility one has to compare it to the capabilities of
dedicated second generation $b$
physics experiments LHC-b and BTeV at hadronic colliders and the
extension of the first generation experiments BELLE and BaBar
running with significantly increased luminosities of possibly
$10^{35}s^{-1}cm^{-2}$. We refer to the latter as
BELLEII/BaBarII; if it is
realized, GIGA-Z would not  add anything new to most measurements of
$B_d$ and $B_u$ decays (some possible exceptions are stated below).
On the other hand we consider it very unlikely that BELLEII/BaBarII
would ever spend quality time above the $\Upsilon (4S)$ to study
$B_s$, let alone $\Lambda _b$ decays.

The comparison with the potential of LHC-b and BTeV is less
straightforward.
The number of $b$ quarks produced at
GIGA-Z has to be compared with
$10^{12}$ to $10^{13}$ $b$ quarks produced at hadronic collider
experiments \cite{LHC-bTDR}. However this large
number is contained in a enormous background, the signal-to-noise
being typically $S/N \approx 5 \times 10^{-3}$ which has to be compared
to $S/N \approx 0.21$ at GIGA-Z.
 Another advantage of GIGA-Z over LHC-b / BTeV is the efficiency of
the flavour tag, i.e. of discriminating between $b$ and $\bar{b}$.
Using polarized beams at
GIGA-Z will result in substantial forward-backward asymmetry
for the $b$ quarks which will make flavour tagging relatively easy.
Typically one can expect an efficiency $\times$ purity of about 60\% at
GIGA-Z due to the
forward backward asymmetry and the cleaner environment, while the
corresponding number at the hadronic fixed target experiments is
typically a factor of ten lower, i.e. about 6\%. A more detailed discussion
can be found in \cite{HawkMoe}.

In both GIGA-Z as well as at LHC-b / BTeV all
beauty hadrons are accessible. In order to estimate the
production rate for each species one may use the numbers obtained
at LEP or theoretical estimates. The relevant numbers are listed
in table~\ref{tab1}.

\begin{table}[h]
\begin{center}
\begin{tabular}{|l|l||l|l|}
\hline
mode & $b$ branching ratio & mode & $b$ branching ratio \\
\hline
$b \to B_u$ & 40\% & $b \to B_d$ & 40\% \\
\hline
$b \to B_s$ & 10\% & $b \to \Lambda_b$ & 10\% \\
\hline
$b \to B^{**}$ & $\approx$ 25\% & $b \to B_c$ & $\approx 10^{-3} -
10^{-4}$ \\
\hline
$b \to (bcq)$ & $\sim 10^{-5}$ & & \\
\hline
\end{tabular}
\end{center}
\caption{$b$ quark branching fractions, taken from \protect{\cite{PDG2000}},
except the ones for $B_c$ and $(bcq)$
which are from \protect{\cite{Rueckl}}.}
\label{tab1}
\end{table}

Precise data on $CP$ violation are one of the
major goals at the presently running $B$ factories and of
next generation
$b$ physics experiments at hadron machines. Some of these experiments
are designed to provide a good measurement of $CP$ asymmetries in
$B$ decays and thus it is very likely that at the time of GIGA-Z
$CP$ violation will be well studied. Even with the
advantage in tagging efficiency and signal-to-noise ratio at GIGA-Z
the huge statistics at the hadronic collider experiments
will win, at least in the standard modes used for the $CP$ violation
analyses. We shall give
a summary of the relevant numbers in the next subsection.

Yet we want to point out that the GIGA-Z option
would provide a truly complementary program by addressing
three topics within $b$ physics that
are of fundamental importance for the comprehensive
$b$ physics program that is being undertaken worldwide,
yet cannot be addressed adequately at other existing or planned
facilities and therefore will be highly topical even in 2010.

Those topics are:
\begin{enumerate}
\item
Polarized beams will produce a huge sample of highly polarized
beauty baryons whose weak decays can be analyzed. In this way a
determination of the handedness of a quark transition becomes feasible.
\item
The quark level transition
\begin{equation}
b \to q + \nu \bar \nu
\label{bqnunu}
\end{equation}
could well be affected significantly by New Physics in ways
quite different from
$b \to q + l^+l^-$ \cite{GrosLigNar}.
Searching for $b \to q \nu \bar \nu$ in hadronic colliders appears
a hopeless enterprise, and even for a $\Upsilon (4S)$ experiment
it poses quite a challenge.
\item
With the $CP$ asymmetries being functions of the moduli of the
CKM parameters  one attempts to extract
the latter from $CP$ insensitive rates as precisely as possible
to infer the size of the former. $|V(cb)|$ and $|V(ub)|$
are determined in semileptonic $B$ decays. Yet
there is one source of potentially
considerable uncertainties in the values
thus obtained, namely limitations to the validity of quark-hadron
duality, of which at present little is known for certain. Detailed
comparisons of semileptonic
$B_s$ and $B_{u,d}$ decays  would be of invaluable help in this respect.
\end{enumerate}
In the following subsections we shall elaborate on the
points raised above.

\section{Standard Model $CP$ Violation}
The first task is to measure the three angles that
are usually referred to
as
$\phi _1 = \beta$, $\phi _2 = \alpha$ and $\phi _3 = \gamma$
\cite{PDG2000}.

BELLE and BaBar expect to measure sin$(2\beta )$ from the
$CP$ asymmetry in the `golden' mode
$B_d \to \psi K_S$ (and related ones) with an uncertainty of
about 8\%; it should be reduced significantly by
BELLEII/BaBarII. Since this mode is relatively easy to detect
even with large
backgrounds, the high statistics at LHC-b and BTeV
are very likely to win
over GIGA-Z. Indeed, the precision achieved at LHC-b
is expected to be $\sigma(\sin 2\beta)\approx 1.5\%$, and similar
for BTeV.

The angle $\alpha$ can be determined from $CP$ asymmetries
in $B\to \pi 's$. However the observable asymmetries
are significantly or even severely affected by hadronic
uncertainties. One way which has been studied in some detail is
to measure the branching ratio for all $B\to \pi \pi$ modes together
with the asymmetry in $B_d \to \pi^+\pi^-$: $\alpha$
can then be obtained through an isospin analyis of the two pion final
state \cite{gronau}. Yet the channel
$B_d \to \pi^0 \pi^0$ is estimated to have a tiny branching ratio;
furthermore, it is very hard to identify at LHC-b / BTeV. In \cite{LHC-bTDR}
the prospects of extracting $\alpha$ at LHC-b have been considered on  
the basis of $B \to \pi^+ \pi^-$ only, assuming that the penguin  
contribution to these decays are known; based on this
an uncertainty $\sigma(\alpha) \sim 3^\circ - 10^\circ$,
(depending on the value of $\alpha$) could be achieved.
BTeV on the other hand will rely on a
detailed analysis of the Dalitz plot for $B\to 3 \pi$.
The cleaner environment for both BELLEII/BaBarII as well as
GIGA-Z should make the measurement of
$B_d \to\pi^0\pi^0$ easier. For the case of GIGA-Z a first
discussion on this can be found in \cite{HawkMoe}; however,
a detailed study of $B_d \to\pi^0\pi^0$ at GIGA-Z
has not yet been performd.

Determining $\gamma$ at the first generation experiments will be extremely
difficult, since the $B_s$ states are not accessible.
It is expected that at LHC-b/BTeV
one can  determine $\gamma$ through a combination of methods with an
uncertainty of $\sigma(\gamma) \sim 6^\circ - 14^\circ$
depending on  the $B_s$ mixing parameter and strong phases.
Furthermore the Cabibbo suppressed angle often referred to
as $\chi$ can be measured through the $CP$ asymmetry in
$B_s \to \psi \eta , \psi \phi$. A signal beyond the expected value
for the asymmetry of about 2\% would reveal the presence of New
Physics. The sensitivity of LHC-b/BTeV should reach the CKM level.
With most of the relevant modes suffering from small branching ratios,
yet possessing clear signatures, we do not see how GIGA-Z could be
competitive with LHC-b/BTeV.

There are actually six KM unitarity triangles rather than one with
angles of order unity, $\lambda^2$ and $\lambda ^4$. Ultimately
one wants to measure as many of them as possible; yet again no case
can be made that GIGA-Z could in general overcome its intrinsic
disadvantage  in statistics relative to LHC-b/BTeV.

\section{Weak Decays of Polarized Beauty Baryons}

\begin{table}
\begin{center}
\begin{tabular}{|l|c|c|}
\hline
Mode & branching ratio & number of events \\
\hline
\hline
$\Lambda_b \to \Lambda_c \ell \bar{\nu}_\ell$
           & $8 \times 10^{-2}$ & $4.7 \times 10^{6}$\\
\hline
$\Lambda_b \to p \ell \bar{\nu}_\ell$
           & $8 \times 10^{-4}$ & $4.7 \times 10^{4}$\\
\hline
$\Lambda_b \to X_s \gamma$ & $2.7 \times 10^{-4}$ & $11 000$ \\
\hline
$\Lambda_b \to \Lambda \gamma$ &  $3.7 \times 10^{-5}$ & $ 1400 $ \\
\hline
$\Lambda_b \to \Lambda \ell \ell$ & $ 1.2 \times 10^{-6}$ & 50 \\
\hline
\end{tabular}
\end{center}
\caption{Expected numbers of events for $\Lambda_b$ decays, based
         on the standard model estimates}
\label{tab:2}
\end{table}

A fundamentally unique feature of GIGA-Z would be the availability
of polarized beams. They would yield polarized beauty quarks which
in turn give rise to highly polarized beauty baryons through
fragmentation in about 10\% of the events which still represents a
huge number. This opens up
a whole new field of dynamical  information one can deduce from
their weak
decays. The  existence of initial state polarization in
$\Lambda _b$ decays allows to analyze the chirality of the
quark coupling {\em directly}; it also leads to
a whole new program of studying
observables revealing direct $CP$ violation.

In analysing $b \to s \gamma$ in a generic way one has to
keep in mind that there are actually two transition operators, namely
\begin{equation}
b_R \to s_L \gamma \; , \; b_L \to s_R \gamma \; .
\end{equation}
While the second one is highly suppressed in the SM
(by ${\cal O}(m_s/m_b)$) they could be of comparable size
in New Physics scenarios:
\begin{eqnarray} \label{SM}
{\rm Standard\; Model:} \; \; \;
b_R \to q_L \; \; \; &\gg& \; \; \;
b_L \to q_R \\
{\rm New \; Physics:} \; \; \;
b_R \to q_L \; \; \; &\sim& \; \; \;
b_L \to q_R
\end{eqnarray}
While the decays of mesons realistically
cannot distinguish between these two transitions, those
of baryons can. An study of the $\Lambda$ polarization in
the decay $\Lambda_b \to \Lambda \gamma$ with polarized $\Lambda_b$
would probe the SM prediction (\ref{SM}).
Such a study should be quite feasible with the expected sample size,
see Table (\ref{tab:2}). A significant non-vanishing contribution of
$b_L \to s_R \gamma$ would signal the
intervention of New Physics. One can actually undertake an
{\em inclusive} polarization study of $\Lambda _b \to \Lambda  \gamma +X$
with large statistics; the clean environment of GIGA-Z is of course
crucial here.
Corresponding studies can be performed with
$\Lambda _b\to l^+l^-X$ with smaller statistics.

Semileptonic decays of polarized $\Lambda _b$ allow to test
the $V-A$ character of $b$ quarks with unprecedented accuracy.

The next step is to search for $CP$ asymmetries in the
{\em spectra} of semileptonic decays. Since such asymmetries
require the intervention of New Physics, one has a better
chance in $b\to u$ than $b\to c$ transitions and in those that
involve the suppressed chirality state; i.e., one would
compare
\beq
\Lambda _b \to l^- (p+X)_{\rm no\; charm}
\; \; \; vs. \; \; \;
\bar \Lambda _b \to l^+(\bar p+X)_{\rm no\; charm}
\eeq
for polarized $\Lambda_b$'s.

In final states with at least three particles
-- $\Lambda _b \to ABC$ -- one can also form  T-odd correlations like
\begin{equation}
C_T \equiv \langle \vec \sigma _{\Lambda _b} \cdot
(\vec p_A \times \vec p_B)\rangle
\end{equation}
with $\vec p_{A[B]}$ denoting the momenta of $A$ and $B$, respectively,
and $\vec \sigma _{\Lambda _b}$ the $\Lambda _b$ polarization.
$C_T \neq 0$ can be due to T violation -- or final state interactions.
This ambiguity can be resolved by observing the $CP$ conjugate
process $\bar \Lambda _b \to \bar A \bar B \bar C$ and the
analoguous observable $\bar C_T$: if one finds
\begin{equation}
C_T \neq \bar C_T \; ,
\end{equation}
then one has uncovered $CP$ violation of the direct variety.
Since these effects are typically quite suppressed in the
Standard Model, such studies represent largely a search for
New Physics. They can be performed in nonleptonic modes
\begin{equation}
\Lambda ^0_b \to \Lambda ^+_c \pi ^- \pi ^0, p \pi ^- \pi ^0,
\Lambda K^+ \pi ^-
\end{equation}
as well as in semileptonic channels containing a
$\tau$ lepton, since the effect is proportional to the
lepton mass \cite{BENSON}:
\begin{equation}
\Lambda ^0_b \to \Lambda ^+_c \tau ^- \nu, p \tau ^- \nu
\end{equation}

In passing we would like to note that analogous studies can be
performed with polarized {\em charm} baryons.

\section{$B \to X_q \nu \bar \nu$}

A measurement of this mode is impossible at the hadronic machines due
to large backgrounds. Aside from the cleaner environment,
a $Z^0$ factory has one important intrinsic advantage here, namely
that the $b$ quarks are produced into different hemispheres. This
hemispheric separation and the simplicity of the underlying
event would provide powerful tools in searching and actually
measuring such transitions. This is also illustrated by the
fact that the present bound was deduced at LEPI:
\begin{equation}
{\rm BR}(B \to X_s \nu \bar \nu) \leq 7.7 \cdot 10^{-4}
\; \; \; {\rm ALEPH}
\end{equation}
In table~\ref{tab:3} we collect the standard model expectations
\cite{GrosLigNar,ALI}
for decays of the type $b \to s  \nu \bar{\nu}$.
\begin{table}
\begin{center}
\begin{tabular}{|l|c|c|}
\hline
Mode & branching ratio & number of events \\
\hline
\hline
$B \to X_s \nu \bar{\nu} $ & $4 \times 10^{-5}$ &  $19000$  \\
\hline
$B \to K \nu \bar{\nu}$ &$ 2.4  \times 10^{-6}$ & $1150 $  \\
\hline
$B \to K^* \nu \bar{\nu}$ &$5 \times 10^{-6}$  & $2400 $ \\
\hline
\end{tabular}
\end{center}
\caption{Expected numbers of events for $b \to s \nu \bar{\nu}$
         decays, based on standard model estimates. A sum over the
         neutrino species is understood. The numbers are from
         \protect{\cite{BB93}} for the inclusive and from
         \protect{\cite{CFSS}} for the exclusive decays.}
\label{tab:3}
\end{table}

New Physics can affect $b \to q l^+l^-$ and
$b \to q \nu \bar \nu$ in quite different ways for various
reasons \cite{ALI}.
Of course, $b\to q \nu \bar \nu$ provides
hardly a spectacular signature. Therefore searching for
such modes in the environment of a hadronic collider appears
quite hopeless.  In an $e^+e^-$ threshold machine such
transitions could be found only at the cost of reconstructing
one $B$ more or less fully.

At GIGA-Z  the statistics will be high enough to  make
such searches a meaningful enterprise. As can be seen from
table~\ref{tab:2} one can expect a few times $10^3$ events in exclusive
channels and about $10^4$ inclusively, based on the standard model
rates.

\section{Semileptonic $B_s$ Decays}

One of the motivations for extracting reliable values for
the CKM parameters $|V(cb)|$, $|V(ub)|$ etc. is to infer
predictions for various $CP$ asymmetries in $B$ decays that
are as precise as possible. $|V(cb)|$ and $|V(ub)|$ are
determined in semileptonic $B$ decays through observables
in exclusive as well as inclusive modes.

The most reliable
values for $|V(cb)|$ have been derived from the total
semileptonic width of $B$ mesons and from the rate of
$B\to l \nu D^*$ at zero recoil, and the two values agree
in a nontrivial fashion; the theoretical uncertainties
are estimated to be around $5\%$. They might be reduced
for $|V(cb)|$ from $\Gamma _{SL}(B)$ to about 2\% with
similar progress concerning $B\to l \nu D^*$ less
likely.

The
situation is much less  satisfactory with respect to $|V(ub)|$: its
determinations so far have a sizeable model dependence with little
theoretical  control. There is thus little guidance in the
theoretical error estimate, but a realistic estimate assuming some
theoretical progress is a theoretical uncertainty of about
$(10-15)\%$.

Certainly after BELLEI/BaBarI -- if not before -- the theoretical
treatment will become the limiting factor.

However on top of that there could conceivably be another
significant source of a systematic uncertainty: quark-hadron
duality or duality for short which underlies almost all
applications of the $1/m_Q$ expansions cannot be an identity.
There is a large
body of folkloric or circumstantial evidence that duality is a
useful and meaningful concept in particular for
semileptonic transitions. Yet for a full evaluation of the
comprehensive data on beauty physics as they will exist in
2010 it is essential to know with {\em tested} confidence whether
limitations to duality in semileptonic
transitions arise on the 10\%, the 5\% or the 1\% level.
It is quite unlikely that this question can be answered
by theoretical means alone. A much more realistic way is
to proceed like one does in dealing with experimental
systematic uncertainties: undertake to extract the same
quantity in systematically different way and compare
the results.

More specifically one can perform an ``independent''
extraction of
$|V_{cb}|$ in $B_s$ decays.
This could be done through measuring $\Gamma _{SL}(B_s)$.
Secondly, one could
determine the rate for $B_s \to l \nu D^*_s $, extrapolate
to zero recoil and extract $|V(cb)F_{B_s \to D_s^*}(0)|$. Note that
the Heavy Quark Expansion yields
\begin{equation}
|F_{B_s \to D_s^*}(0)| \simeq |F_{B \to D^*}(0)|
\end{equation}
up to $SU(3)$ breaking corrections, which can be estimated.

Extracting $|V(cb)|$ separately from $B_d$ and $B_u$ decays is an
important cross check on experimental systematic uncertainties, but
in all likelihood it could {\em not} reveal duality violations.
For the physical origin of those would be the
`accidental' presence of a nearby
hadronic resonance with appropriate quantum numbers; up to small
isopin breakings it would affect equally
$B_d \to l \nu X_c$ and $B_u \to l \nu X_c$, but not
$B_s \to l \nu X_c$; likewise a resonance affecting $B_s$
transitions would have no impact on $B_{u,d}$ channels.
If the same value emerged for $|V(cb)|$ in both cases, we would
truly have
established theoretical control in this case at least. If not, we
would not know which of the values is the correct one, but
we would be aware of a serious problem.

Duality violation could exhibit a different pattern in
$B\to l \nu X_u$ channels. Here a detailed comparison of
$B_d$ and $B_u$ modes is called for also theoretically:
on one hand one expects a difference in the endpoint region of
$B_d$ and $B_u$ semileptonic decays \cite{WA}, and on the other
hand hadronic resonances could affect $B_d\to l \nu X_u$ and
$B_u \to l \nu X_u$ quite differently. Yet even so,
$B_s \to l \nu X_u$, both inclusively and exclusively,
would provide crucial cross checks.

\section{Summary}

A GIGA-Z facility could not be realized before 2010 at the
earliest. We certainly hope and fully anticipate that the usually
discussed  $CP$ studies in the beauty hadron sector will have been
performed with considerable accuracy by that time or soon thereafter.
Observing the same reactions in the GIGA-Z environment might turn
out to serve some purpose still -- but that is {\em not} the
motivation we are suggesting in this note.

Our emphasis is on
GIGA-Z providing us with novel and unique tools to probe for the
presence of New Physics in the beauty sector: on one hand one can
study the weak decays of polarized beauty baryons and
$B\to \nu \bar \nu X$ transitions; on the other hand one can
perform detailed analyses of in particular $B_s$ decays to
cross check systematic uncertainties in our
extractions of $|V(cb)|$ and
$|V(ub)|$ which form the basis of the CKM predictions for the
asymmetries already observed.


\end{document}